\documentclass[prb,aps,preprint,showpacs]{revtex4-1}
\def\alt{\ {\raise -0.25em
\hbox{{$\buildrel < \over \sim$}}}\ } \def\agt{\ {\raise -0.25em
\hbox{{$\buildrel > \over \sim$}}}\ }

\usepackage{graphicx}
\usepackage{amsmath}
\usepackage{bm}
\usepackage{amsbsy}

\begin{document}
\title{Magnetic Transport in Spin Antiferromagnets 
for Spintronics Applications}
\author{Mohamed Azzouz} 
\email[E-mails to be sent to: ]{mazzouz@laurentian.ca}
\affiliation{Department of Physics, Laurentian University, 
Ramsey Lake Road, Sudbury, Ontario, Canada P3E 2C6}
%
\begin{abstract} 
Had magnetic monopoles been ubiquitous as electrons are, we would 
probably have had
a different form of matter, and
power plants based on currents of these magnetic charges would have been
a familiar scene of modern technology.
Magnetic dipoles do exist, however, and in principle 
one could wonder if we can use them to generate
magnetic currents.
In the present work, we address
the issue of generating magnetic currents and magnetic
thermal currents  
in electrically-insulating low-dimensional Heisenberg  
antiferromagnets by invoking
the (broken) electricity-magnetism 
duality symmetry. The ground state of
these materials is a spin-liquid state 
that can be described well via~the Jordan--Wigner fermions, which
permit an easy definition of the magnetic particle and thermal
currents. The magnetic and magnetic thermal
conductivities are calculated in the present 
work using the bond--mean field theory.
The spin-liquid states in these antiferromagnets
are either gapless or gapped liquids of 
spinless fermions whose flow defines a current 
just as the one defined for electrons in a Fermi liquid. 
The driving force for the magnetic current
is a magnetic field with a gradient along the magnetic 
conductor. We predict the generation of a magneto-motive
force and realization of magnetic circuits using low-dimensional 
Heisenberg antiferromagnets. 
The present work is also about claiming that
what the experiments in spintronics attempt to do 
is trying to treat the magnetic degrees of freedoms on the same footing
as the electronic ones. 
\end{abstract}

\maketitle

\section{Introduction}

The issue of the adequate definition of 
the spin current had attracted significant interest
because of its importance in 
spintronics' applications \cite{gregg2001,jedema2001,sharma2005}.
An {et al}.
\cite{an2012} 
used the relativistic Dirac equation in 
order to define such a current. In addition, many other authors argued that 
the spin transport includes both linear displacement of 
spins as well as angular motion due to the rotation of the spins.
One~of the earliest 
problems encountered in the definition of the 
spin current is the
satisfaction of the continuity equation \cite{sun2005,choi2014}.
It is interesting that spintronics
experiments attempt to marry in practice between 
spin currents and electric currents,
and create one current from the other and vice versa.
It is as if these experiments try
to prove in a practical manner some 
sort of symmetry between 
electricity and magnetism. We propose that, 
at a more fundamental
level, these experiments attempt to prove the duality 
electricity--magnetism symmetry, which is missing from
the Maxwell equations in the presence of matter 
(Maxwell equations are symmetric under the duality 
symmetry transformation
in vacuum).
While this symmetry is broken at the monopole level, 
it could approximately hold at the dipole level in 
materials where the charge
degrees of freedom are practically frozen due to 
a large energy gap in
their excitations.
In such materials, the 
magnetic degrees of freedom 
carried by magnetic dipoles
are responsible for the low-lying energy excitations.
The low-dimensional (chains and ladders) Heisenberg
antiferromagnets constitute a good example of these materials.

For these low-dimensional antiferromagnets, 
a natural way to deal with the difficulties associated
with the definition of the magnetic particle and heat 
currents
is the usage of the duality symmetry of electromagnetism.
In the remainder of this manuscript, we will refer to the spin current 
and spin thermal current in these materials as magnetic current 
and magnetic thermal current, respectively.
This relabeling is necessary in order to best reflect 
this symmetry between electricity and magnetism.
It is well known that the 
Maxwell equations would have been fully
symmetric under the duality transformation 
if magnetic monopoles existed.
If they did, magnetic currents would have been 
defined in the same way as 
electric currents. In the present real situation where 
the duality symmetry between electricity and magnetism is broken
in the presence of sources (matter),
the magnetic 
dipoles resulting from the spins' degrees of freedom 
of electrons do exist, however.
In the Heisenberg antiferromagnets,
these magnetic dipoles interact and form the so-called spin liquids
that bear interesting similarities with the Fermi liquid states of
electrons in conventional metals.

We define the magnetic current and magnetic thermal current
after transforming
the spin degrees of freedom using the Jordan--Wigner 
(JW) transformation in one-dimension (1D), 
or its generalized sisters in the case of
ladders \cite{azzouz1993,azzouz2001}.
This approach is well suited for insulating antiferromagnets,
like the linear-chain compound Sr$_2$CuO$_3$. Such materials 
are electrically insulating because 
the electric charge degrees of freedom 
are suppressed by large excitation energy gaps,
and are characterized by spin-$1/2$ moments that are arranged 
on chains or ladders.
Due to their strong spatial anisotropic magnetic 
exchange interactions and large
quantum fluctuations, they do not magnetically
order even at 
very low temperatures. One of the interesting consequences of using the
JW transformation is the definition 
of a magnetic current (rather than a spin current) because 
such a transformation puts the treatment of the spin degrees 
of freedom on the same footing as the electronic charge degrees 
of freedom in metals.  
We claim that this one-to-one correspondence between
the magnetic moments (spins) in the Heisenberg 
antiferromagnets and electrons (charges) in metals
is reminiscent of the duality symmetry
of electricity and magnetism in vacuum, or even 
in matter had the magnetic monopoles \cite{goddard1978} 
been ubiquitous as electrons do--that is to say, that 
the original 1D JW and its
higher dimension generalized sisters
transform the spins into spinless fermions that behave exactly 
like electrons as far as Fermi statistics 
and transport properties are concerned.

The Heisenberg quantum antiferromagnets are modeled with
the Heisenberg Hamiltonian that consists of exchange interactions
between spins on adjacent sites. The 1D case relevant for 
Sr$_2$CuO$_3$, for example, is simpler 
to analyze, and will be used throughout this paper. Note, however, that
the results of this work can be generalized to 
three-leg ladder systems, which
behave as effective single Heisenberg chains
especially when the interchain interaction is much greater than the intrachain 
one~\cite{azzouz2005}.
The~two-leg Heisenberg ladder is, however, gapped~\cite{azzouz2006},
and an approach will be developed in the near future 
by taking into account this gap.
Upon using the JW transformation, 
the 1D Hamiltonian maps into that of spinless fermions with 
a tight-binding kinetic energy term corresponding 
to the XY part of the spin Hamiltonian, and a repulsive interaction between
JW fermions on adjacent sites resulting from the Ising term of the 
Hamiltonian. Afterwards, we define particle and thermal
currents for these spinless fermions
in the same way as for electrons in a metal, and use 
the techniques of transport theory
including the Kubo formula for 
calculating the conductivity and the Green--Kubo formula for 
the magnetic thermal conductivity. 

The driving force for the magnetic current of the 
spinless fermions can be provided by an external magnetic 
field with a gradient along the chain of spins. 
The reason for this is that the JW transformation
maps the magnetic field in the Zeeman-coupling term onto 
the chemical potential for the spinless JW fermions, as is well known.
Thus, a gradient in the magnetic field forces the spinless fermions to flow
along the chain in order to lower their energy, just as electrons do
in order to lower their energy when a gradient in the chemical potential 
is applied to them. Note that 
a magnetic field with a gradient, rather~than a uniform 
magnetic field alone, is needed for the present 
case of a magnetic current
because this magnetic current is not 
that of magnetic monopoles, but that
of magnetic dipoles. This is similar to the fact that 
a gradient in the electric field can be the driving force
for an electric dipole.
The~experimental work by Hirobe {et al.} \cite{hirobe2016}
reported the observation of spin current in Sr$_2$CuO$_3$, which~resulted from a temperature gradient. Indeed
a temperature gradient $\nabla T$ 
can generate a flow of the JW fermions
just as it does for electrons in metals.
We, however, argue for and support
the more convenient utilization of a magnetic field gradient.
It is worth mentioning that
we think that the Heisenberg antiferromagnets 
can be incorporated into spintronics devices
without using electric contacts. The magnetic fields generated by 
circulating electric currents in the regular electric circuits 
of a given device can 
be taken advantage of to induce a magnetic current in the 
Heisenberg antiferromagnet part of the device.

The present paper is organized as follows. 
In Section \ref{section2},
the nature of the JW fermions is discussed in connection
with the (broken) electricity-magnetism duality
symmetry. In
Section \ref{section3},
a review of the bond--mean-field 
theory (BMFT) applied to the Heisenberg chain 
in a magnetic field is presented. 
In Section \ref{section4}, 
the particle current density, 
the Green's and spectral functions are calculated for the JW fermions.
In Section \ref{section5}, 
the current--current correlation function 
is calculated and used to derive the conductivity of the JW fermions.
Section \ref{section6} deals with the calculation 
of the magnetic thermal conductivity.
In Section \ref{section7}, the main result of the present work 
is discussed, and predictions for potential applications
are outlined.
Conclusions are drawn in Section \ref{section8}.

\section{The JW Transformation and Duality Symmetry}
\label{section2}

In addition to the fact that the JW transformation
preserves the spin commutation relation as required, we think
that a more profound aspect of this transformation is 
related to the 
electricity--magnetism duality (broken) symmetry
as explained in the introduction.
This transformation maps magnetic dipoles (magnetic degrees
of freedom)
onto spinless fermions whose statistics and physics are
similar to the ones of electrons, 
except for the electric charge.
The 1D JW transformation~\cite{jordan1928} reads as:
\begin{equation} 
\begin{array} {lll}
S_i^-&=&c_ie^{i\sum_{j=1}^{i-1}c_j^\dag c_j}, \cr
S_i^z&=&c_i^\dag c_i -1/2,
\label{1DJW}
\end{array}
\end{equation}
where $i$ or $j$ label the chain sites, $S_i^-$ is
the spin ladder operator, and $S_i^z$ is the $z$-component
of the spin operator. $c_i$ ($c_i^\dag$) is 
the JW annihilation (creation)
operator.
We believe that there is a profound reason behind 
the fact that the JW fermions satisfy the same statistics
as the original electrons that carry the spin (thus magnetic) degrees 
of freedom, and that this is not a mere accident. 
The JW transformation bears in it the footprint 
of the electricity-magnetism
duality symmetry of the Maxwell equations in the vacuum. 
The rational for this proposal is that
if the Maxwell equations included magnetic monopoles,
one could have had a transformation between electrons and magnetic 
monopoles in any given study of the electronic and magnetic
properties of any material, and that 
the magnetic and electronic properties would have been transformed 
naturally into each other as a consequence of this duality symmetry.
For~example, we could have had defined easily the magnetic current 
of magnetic monopoles. However, since the duality symmetry does not 
apply in the presence of matter, 
the magnetic degrees 
of freedom, which derive
their meaning only from the electronic ones (electrons' spin here), 
are transformed into the JW 
fermions for the Heisenberg antiferromagnets.
This transformation tells us that
the magnetic dipoles are fermions that do not carry a spin.
In the next section, we will use this symmetry to predict and argue
that it is possible to construct magnetic generators and 
magnetic circuits made of Heisenberg antiferromagnets.
This constitutes the central finding of the present work.

\section{The Magnetic Current in the Bond--Mean Field Theory}
\label{section3}

In the presence of 
a magnetic field gradient along the  
chain, the Heisenberg model assumes the~form:
\begin{eqnarray}
H_{1D}=J\sum_{i}{\bf S}_{i}\cdot{\bf S}_{i+1} - 
g\mu_B\sum_i B^{\rm ex}_iS_i^z,
\label{hamiltonian 1D model}
\end{eqnarray}
where $J$ is the exchange 
coupling constant and
$B^{\rm ex}_i$ is a position dependent magnetic field that can be 
taken to vary linearly 
with position along the chain; i.e., 
$B^{\rm ex}_i=B_0^{\rm ex}x$ with $B_0^{\rm ex}$  a field per unit length.
In~terms of the JW fermions, Equation (\ref{1DJW}), this Hamiltonian maps onto:
\begin{eqnarray}
H_{1D}=\frac{J}{2}\sum_{i}{c}_{i}^\dag{c}_{i+1} + {\rm H.C.} +
J\sum_i \big(c_i^\dag c_i-\frac{1}{2}\big)
\big(c_{i+1}^\dag c_{i+1}-\frac{1}{2}\big) 
- \sum_i h_i c_i^\dag c_i +\sum_i \frac{h_i}{2},
\label{hamiltonian 1D model JW}
\end{eqnarray}
where $h_i=g\mu_B B^{\rm ex}_i$, 
with $g$ being the Land\'e factor
and $\mu_B$ the Bohr magneton. As is well known, 
a~constant magnetic field $h_i\equiv h$ 
plays the role of the chemical potential
for the JW fermions. Such~a~constant magnetic field
is also known to polarize the spins along its direction, with the 
magnetization $M_z=\chi h$ for $h\ll J$, where $\chi$ is
the uniform spin susceptibility.
A gradient in the magnetic field along the chain is equivalent to tilting 
the chemical potential. Such tilting causes the JW fermions 
to flow, thus creating a current of these fermionic particles. 
For the spin degrees of freedom,
the flow of the JW fermions occurs with hopping amplitude
$J$/2 and results in the flow of spin flip fluctuations,
since the presence of a JW fermion at a given site 
is a spin up and its absence is identified with a spin down; 
Equation (\ref{1DJW}).

The need for a gradient in the magnetic field to drag
the magnetic excitations resulting from the magnetic dipoles
is similar to the 
fact that a gradient in an electric field is needed in order
to drag electric dipoles; a uniform electric field alone
does not act on the dipoles, except by a force couple. 
We~think that this similarity 
is a consequence of the 
duality symmetry between electricity and magnetism.
This, in turn, supports our claim that
the JW transformation bears a
signature of this~symmetry.

We assume that 
the gradient in the magnetic field
is much smaller than the spin exchange coupling.
We thus calculate the magnetic conductivity and magnetic thermal 
conductivity using the (Green--)Kubo formula
in the linear response approximation. The current of the JW fermions
is readily defined as the magnetic current, and
the current--current
correlation function is evaluated in the limit of 
a uniform magnetic field to get this conductivity.


In Ref. \cite{azzouz2006}, the effect of a 
uniform field on the Heisenberg chain
was investigated in the framework of the BMFT \cite{azzouz1993}. In brief,
the Hamiltonian in the presence of the Zeeman coupling term
$g\mu_BB\sum_i S_i^z$ with a uniform magnetic field $B$
along the $z$-axis, takes on the form:
\begin{eqnarray}
H_{1D}
= NJQ^2 + Nh/2 - NM_z(M_z+1)J + \sum_k \Psi^{\dag}_{ k} 
{\cal H}_{1D}(h)\Psi_{ k},
\label{kham1dH}
\end{eqnarray}
where $N$ is the total number of sites, and 
$M_z = \langle S_i^z\rangle$ is the magnetization per site.
The two-component 
spinor $\Psi_k$ is given by:
\begin{eqnarray}
%
\Psi_k=
\left(\begin{array}{c}
		  c^A_{k}\\
		  c^B_{k}
\end{array}
\right),
\label{spinor}
\end{eqnarray}
and
the Hamiltonian density matrix by:
\begin{eqnarray}
{\cal H}_{1D}(h)= (2M_zJ-h)\sigma_0 - (J_1\sin k)\sigma_2,
\label{hamiltoniandensityH1D}
\end{eqnarray}
where $\sigma_0$ is the $2\times2$ identity matrix 
and $\sigma_2$
the second Pauli matrix.
Here,
$J_1=J(1+2Q)$ with
$Q=|\langle c_ic^\dag_{i+1}\rangle|$ having been 
defined as the spin bond parameter,
and $h=g\mu_BB$. 
The chain is subdivided into two sublattices $A$ and $B$ as
a consequence of the strong antiferromagnetic correlations
that decay only algebraically with distance because 
the ground--state correlation length is infinite. Locally, 
the~spins maintain a staggered orientation that 
justifies the use of the bipartite character.
This gives 
rise to two types of JW fermions at the mean-field level, and the 
creation and annihilation operators are labeled by the
two sublattice indexes.
At the mean-field level,  
the chemical potential renormalizes to 
$h'=h-2M_zJ=h(1-2J\chi)$ if $h\ll J$, or 
to $h'=h-J$ if $h>h_c$ in the fully saturated state. 
$\chi$~is the uniform spin susceptibility,
and $h_c=2J$ is the magnetic field
above which the magnetization saturates~\cite{bonner1964,parkinson1985}.

Diagonalizing Equation (\ref{hamiltoniandensityH1D}) yields the following 
energy eigenvalues:
$$
E_p(k) = -h' + p J_1|\sin k|;\ \ \ p=\pm.
$$

The magnetization per site, $M_z$, 
and the bond parameter, $Q$, are given by \cite{azzouz2006}:
\begin{equation} 
\begin{array} {lll}
Q &=& -\frac{1}{2}\int \frac{dk}{2\pi} |\sin k|
\sum_{p=\pm}pn_F[E_{p}(k)], \cr
M_z &=& \frac{1}{2}\int\frac{{\rm d}k}{2\pi}\sum_{p=\pm} n_F[E_p(k)] 
- \frac{1}{2},
\label{magnetization1D}
\end{array}
\end{equation}
%
%
%
where $n_F(x)=1/(1+e^{\beta x})$ is the Fermi factor. 
Here, $\beta=1/k_BT$ is inverse temperature.

In order to calculate the magnetic particle and thermal 
conductivities, we will next 
calculate the current density, Green's function and 
spectral function for the JW fermions within the BMFT. 
We deal first with the magnetic~conductivity.

\section{Current Density, Green and Spectral Functions}
\label{section4}
\vspace{-6pt}
\subsection{Current Density Operator}

Because every spin is carried by a localized electron 
on the chain, the spins are not mobile, which means that they cannot create a spin current in the same way as in metals where 
electrons are mobile.
The spin fluctuations can however propagate along the chain, 
thus creating the magnetic current that we seek to calculate.
The spin fluctuations' propagation is caused by the
kinetic energy of the JW fermions in Hamiltonian 
(\ref{hamiltonian 1D model JW}), and this magnetic current's
density operator is therefore given within
the tight-binding approach by: 
\begin{eqnarray}
{\bf j}=-i\frac{J}{2}\sum_i\bigg\{
(c_{2i+1}^{B\dag} c^A_{2i} - c^{A\dag}_{2i}c_{2i+1}^B)
+ (c^{A\dag}_{2i}c_{2i-1}^B - c_{2i-1}^{B\dag} c^A_{2i})
\bigg\} {\hat {\bf x}},
\label{current density}
\end{eqnarray}
where ${\hat {\bf x}}$ is a unit vector along the chain direction.
Using the phase configuration $...\pi-0-\pi...$ on the intersite 
bonds along the chain, 
which is utilized to write the mean-field Hamiltonian in Equation
(\ref{kham1dH}), 
Ref. \cite{azzouz2006}, and 
the Fourier transform
$
c_{i}^{\alpha} = \frac{1}{\sqrt N}\sum_k e^{ikr_i}c_k^\alpha$ where
$\alpha=A,B$,
the operator ${\bf j}$ takes on the form:
\begin{eqnarray}
{\bf j}=\sum_k  i{\bf v}(k)
(c_k^{A\dag} c^B_{k} - c^{B\dag}_{k}c_{k}^A)=\sum_k{\bf v}(k) 
\Psi_k^\dag\sigma_2 \Psi_k,
\label{current density1}
\end{eqnarray}
with ${\bf v}(k)=J\cos k {\hat {\bf x}}$ being the spin velocity 
along the chain. Note the cosine function in this spin velocity
instead of sine because of the above phase configuration.
This spin velocity being in cosine is in agreement with the exact result 
for the energy spectrum $\frac{\pi}{2}J\sin k$, 
Ref. \cite{bonner1964}.

%
%

\subsection{Green and Spectral Functions}

The single-particle Green's function is
defined by ${\cal G}=(i\omega_n - {\cal H}_{1D})^{-1} $. Within the BMFT,
the latter
takes on the $2\times2$ matrix form: 
\begin{eqnarray}
{\cal G}(k,\omega_n)
%
%
=\frac{1}{2} 
\sum_{p=\pm} \frac{\sigma_0-p\sigma_2\sin k/|\sin k|}{i\omega_n-E_p(k)}.
\label{Greens function}
\end{eqnarray}

The retarded Green's function is 
${\cal G}_{ret}(k,\omega)={\cal G}(k,i\omega_n\to\omega+i\eta)$,
with $\eta$ a very small positive~constant.

The spectral function 
${\cal A}(k,\omega)=-2{\rm Im}{\cal G}_{ret}(k,\omega)$ assumes 
the following expression:
\begin{eqnarray}
{\cal A}(k,\omega)=\sum_{p=\pm}
\bigg[\eta\sigma_0+p\bigg(\omega-E_p(k)\bigg)\sigma_2'\frac{\sin k}
{|\sin k|}\bigg]a_p(k,\omega),
\label{spectral function}
\end{eqnarray}
with $a_p(k,\omega)=[\big(\omega-E_p(k)\big)^2+\eta^2]^{-1}$, and 
the matrix $\sigma_2'=-i\sigma_2=\left(\begin{array}{cc}
		  0&-1\\
		  1&0
\end{array}
\right).$

\section{The {\bf j}-{\bf j} Correlation Function and 
Magnetic Conductivity}
\label{section5}

\subsection{Kubo Formula}

The Kubo formula for 
the current-current correlation function, 
$\Pi(q,\tau)=-\frac{1}{V}
\langle  {\bf j}^\dag(q,\tau){\bf j}(q,0)\rangle$, where $V$
is the volume of the sample,
yields in the long-wavelength ($q=0$) limit:
\begin{equation} 
\begin{array} {lll}
\Pi(i\omega_m) &=& \frac{1}{V}\int_0^{\beta}d{\tau}e^{i\omega_m\tau}
\sum_{{\bf k},{\bf k}'}{\bf v}(k'){\bf v}(k) 
\langle
T_{\tau} \Psi_{k'}^\dag(\tau)\sigma_2\Psi_{k'}(\tau) 
\Psi_{k}^\dag(0)\sigma_2\Psi_{k}(0) 
\rangle
\cr
&=& \int_0^{\beta}d{\tau}e^{i\omega_m\tau}
\int_{RBZ} \frac{dk}{2\pi}v^2(k){\rm Tr}
[{\cal G}(k,\tau)\sigma_2{\cal G}(k,-\tau)\sigma_2] \cr
&=&
\int_{RBZ} \frac{dk}{2\pi} \int\frac{d\epsilon}{2\pi}\frac{d\epsilon'}{2\pi}
\frac{n_F(\epsilon)-n_F(\epsilon')}{\epsilon-\epsilon'+i\omega_m}
v^2(k){\rm Tr}[{\cal A}(k,\epsilon')\sigma_2{\cal A}(k,\epsilon)\sigma_2],
\label{j-j correlation function}
\end{array}
\end{equation}
with $\psi_k^1=c_k^A$ and $\psi_k^2=c_k^B$. 
In going from the first line to the second in 
Equation (\ref{j-j correlation function}), the summation over the 3D wavevector 
reduces to only the 1D component, with the summation over the transverse
components, $k_y$ and $k_z$, yielding an overall factor $1/bc$ where $b$ 
and $c$ are the lattice parameters in the $y$~and $z$ directions, respectively.
Note that the sum over the $x$-component of ${\bf k}$, labeled 
$k$, divided by 
the length of the sample $L$ is replaced by 
$\int \frac{dk}{2\pi}$ in the limit 
$L/a\to\infty$ where $a$ is the lattice parameter in the $x$ direction.
For convenience, we set $a=b=c=1$ and will reinstate 
these parameters in the final results of conductivities.
The integral over $k$ in 
Equation (\ref{j-j correlation function})
is carried over the reduced Brillouin zone (RBZ) only 
in order to avoid double counting as a result of the two sublattices. 

\subsection{The Real Part of the Magnetic Conductivity}

We next use the analytical limit $i\omega_n\to\omega + i\eta$
to obtain the retarded conductivity, and write 
$\frac{1}{x+i\eta}=P(x)-i\pi\delta(x)$,
where $P(x)$ is the principal part of $x$ and $\delta(x)$ 
the Dirac delta 
distribution, in~order to cast the conductivity 
in the following form:
\begin{equation} 
\begin{array} {lll}
\sigma(\omega) &=&
\int_{RBZ} \frac{dk}{2\pi} \int\frac{d\epsilon}{4\pi}
\frac{n_F(\epsilon)-n_F(\epsilon+\omega)}{\omega}
v^2(k){\rm Tr}[{\cal A}(k,\epsilon+\omega)\sigma_2{\cal A}(k,\epsilon)\sigma_2] \cr
&=& \int_{RBZ} \frac{dk}{2\pi} \int\frac{d\epsilon}{2\pi} 
\frac{n_F(\epsilon)-n_F(\epsilon+\omega)}{\omega}  v^2(k)
\sum_{p=\pm}\sum_{p'=\pm}\cr
&&\ \ \ \ \ \ \big[\eta^2 - pp'\big(\epsilon+\omega -E_{p'}(k)\big)
\big(\epsilon-E_p(k)\big)\big]
a_{p'}(k,\epsilon+\omega)a_p(k,\epsilon).
\label{conductivity}
\end{array}
\end{equation}

The contribution to $\sigma(\omega)$ in $\eta^2$ gives 
the usual Drude term. The term in $pp'$  
is however negligibly small because 
the main contribution to the integral comes from $\epsilon=E_p(k)$.
Indeed, using~the~representation:
$$
\delta(x)=\frac{1}{\pi}\frac{\eta}{x^2+\eta^2};\ \ \ \eta\to0^+
$$
for the delta distribution, we write 
$
{\eta} a_p(k,\epsilon)\approx \pi\delta\big(\epsilon-E_p(k)\big).
$ 
Then, the conductivity reduces to the semi-classical 
expression \cite{ashcroftmermin}:
\begin{equation} 
\begin{array} {lll}
\sigma(\omega) 
&=& \sum_{p',p=\pm}\int_{RBZ} \frac{dk}{2\pi} \int\frac{d\epsilon}{2} 
\frac{n_F(\epsilon)-n_F(\epsilon+\omega)}{\omega}  v^2(k)\cr
&&\ \ \ \ \ \ \big[\eta - pp'\big(\epsilon+\omega -E_{p'}(k)\big)
\big(\epsilon-E_p(k)\big)/\eta\big]
a_{p'}(k,\epsilon+\omega)\delta\big(\epsilon-E_p(k)\big) \cr
&=& \sum_{p',p=\pm}\int_{RBZ} \frac{dk}{4\pi}  
v^2(k)
\frac{\eta}{[\omega+E_p(k)-E_{p'}(k)]^2+\eta^2} 
\bigg(-\frac{\partial n_F}{\partial \epsilon}\bigg)_{\epsilon=E_p(k)}.
\label{conductivity Drude}
\end{array}
\end{equation}

The direct conductivity 
(DC) is obtained by letting $\omega\to 0$:
%
This gives, after taking into account the doubling
of the unit cell due to the bipartite character 
of the Brillouin zone (BZ),
\begin{eqnarray}
\sigma_{DC}
= \frac{1}{2} 
\sum_{p=\pm}\int_{-\pi}^\pi \frac{dk}{4\pi}  
v^2(k)
\tau\big(E_p(k)\big)
\bigg(-\frac{\partial n_F}{\partial \epsilon}\bigg)_{\epsilon=E_p(k)},
\label{DC conductivity}
\end{eqnarray}
when we assume that the main contribution comes from 
the region of the BZ with $E_p-E_{p'}=0$; i.e.,~near the Fermi energy of the JW fermions since the two bands
$E_+$ and $E_-$ touch at the Fermi energy when the external 
magnetic field is zero.
Here, $\eta^{-1}\equiv\tau\big(E_p(k)\big)$ 
is identified with an energy-dependent
relaxation time. Thus, the magnetic conductivity 
(\ref{conductivity Drude}) and 
DC conductivity (\ref{DC conductivity}) have the same 
form as conductivities of real electrons, with 
the electronic group velocity and energy spectra replaced by those
corresponding to the JW fermions. In principle, one could have 
predicted this result by using the (broken) 
duality symmetry of electricity and magnetism
and stating that the JW transformation is a consequence 
of this symmetry.
For the present case of magnetic currents and conductivities,
the~energy spectra result from the dispersion of the spin fluctuations.
For this reason, it is legitimate to label the current of 
these magnetic dipoles as 
magnetic current because the spins do not move contrary to the
(spintronics) experiments where two opposite currents
of spin-up and spin-down electrons flow.

In the case of a constant relaxation time $\tau_{\rm mag}=1/\eta$,
the integral in Eq. (\ref{DC conductivity}) is simple to evaluate, 
\mbox{and one finds}:
\begin{eqnarray}
\sigma_{DC}\approx 
\frac{\tau J_1a}{\pi\hbar^2bc}=\frac{l_{\rm mag}}{\pi\hbar bc}
\label{DC conductivity near zero T},
\end{eqnarray}
where $l_{\rm mag}=J_1a\tau_{\rm mag}/\hbar$ 
is the magnetic mean-free path. 

The results of the present work tell us that 
all the methods and techniques developed
for the electronic transport can be implemented in the case 
of the Heisenberg antiferromagnets once the JW transformation 
is used to transform the spin degrees of freedom
to spinless fermions. In general,
for any Heisenberg quantum antiferromagnets
in higher dimensions,  
the 2D and 3D JW transformations~\cite{azzouz1993,azzouz2001} can be used,
but one has to deal with the occurrence of long-range 
antiferromagnetic order
below finite critical temperatures for the 3D systems.

\section{The Magnetic Thermal Conductivity}

\label{section6}

Using the definition of the 
the energy current operator ${\bf j}^E$ 
by Zotos, Naef, and Prelovsek \cite{zotos1997}
for the Hamiltonian (\ref{hamiltonian 1D model}), one gets:
\begin{eqnarray}
{\bf j}^E=\frac{J^2}{4}\sum_{i}\big(ic_{i+1}^{\dag}c_{i-1} + {\rm H.c.}\big)
+
\frac{J^2}{2}\sum_{i}\big(-ic_{i+1}^{\dag}c_{i} + {\rm H.c.}\big)
\big(n_{i-1} + n_{i+2} -1\big).
\label{energy current operator}
\end{eqnarray}

In the absence of a magnetic field, $\langle n_i\rangle=1/2$ 
because the magnetization $M_z=\langle n_i\rangle-1/2=0$.
If~we replace $n_i$ by $\langle n_i\rangle$ in 
Equation (\ref{energy current operator}), the energy current density
simplifies to:
\begin{eqnarray}
{\bf j}^E\approx\frac{J^2}{4}\sum_{i}\big(ic_{i+1}^{\dag}c_{i-1} + {\rm H.c.}\big)
+
J^2M_z\sum_{i}\big(-ic_{i+1}^{\dag}c_{i} + {\rm H.c.}\big)
\label{energy current operator2}
\end{eqnarray}
when an external magnetic field is applied along the $z$-axis.
The magnetic thermal current ${\bf j}^Q$ 
is obtained by subtracting $h {\bf j}$,
where $h$ is the chemical potential of the JW fermions, and
${\bf j}=\frac{J}{2}\sum_{i}\big(-ic_{i+1}^{\dag}c_{i} + {\rm H.c.}\big)$ 
is the particle current density. This gives:
\begin{eqnarray}
{\bf j}^Q\approx {\bf j}^E_{\rm MF} - h'{\bf j},
\label{heat current operator3}
\end{eqnarray}
where: 
$$
{\bf j}^E_{\rm MF}=\frac{J^2}{4}
\sum_{i}\big(ic_{i+1}^{\dag}c_{i-1} + {\rm H.c.}\big)
$$
plays the role of the energy current at the mean-field level.
Interestingly, the expression (\ref{heat current operator3}) 
for the magnetic thermal current is the same as
that obtained using the mean-field Hamiltonian with the 
renormalized chemical potential $h'=h-2M_zJ$. 
Taking into account the bipartite 
character of the lattice and transforming into Fourier space yield:
\begin{eqnarray}
{\bf j}^Q\approx\sum_{k} \Psi_k^{\dag}{\cal Q}\Psi_k,
\label{energy current operator4}
\end{eqnarray}
where the $2\times2$ matrix
${\cal Q}= \sum_{i=0}^{1} M_i\sigma_i$
with
$M_0=\frac{J^2}{2}\sin(2k)$ and
$M_1=Jh'\cos k$.
%
%
%
%
In the limit of a weak magnetic field; i.e., $h\ll J$,
which is realized for most
real 1D Heisenberg antiferromagnets, $h'\ll J$ because
$JM_z=J\chi h\ll J$; $\chi\sim 1/J$. In this case, 
${\cal Q}\approx \frac{J^2}{2}\sin(2k)\sigma_0$.
The magnetic thermal conductivity
within the linear response is given by the 
Green--Kubo formalism:
\begin{eqnarray}
\kappa_{\rm mag}=\frac{1}{k_BT^2}\bigg[L^{22} - \frac{(L^{12})^2}{L^{11}}\bigg],
%
%
\label{spin heat conductivity 1}
\end{eqnarray}
with:
\begin{equation} 
\begin{array} {lll}
L^{22}&=&\frac{1}{\omega\beta}
{\rm Im}\int_0^{\beta}
\langle 
{T_\tau{\bf j}^Q}^\dag(\tau){\bf j}^Q(0)
\rangle
d\tau, \cr
L^{11}&=&\frac{1}{\omega\beta}
{\rm Im}\int_0^{\beta}
\langle 
{T_\tau{\bf j}}^\dag(\tau){\bf j}(0)
\rangle
d\tau, \cr
L^{12}&=&\frac{1}{\omega\beta}
{\rm Im}\int_0^{\beta}
\langle 
{T_\tau{\bf j}^Q}^\dag(\tau){\bf j}(0)
\rangle
d\tau.
\label{spin heat conductivity 2}
\end{array}
\end{equation}

We find that $L^{12}=h'L^{11}$; 
$\langle {T_\tau{\bf j}^Q}^\dag(\tau){\bf j}(0)
\rangle= h'
\langle {T_\tau{\bf j}}^\dag(\tau){\bf j}(0)
\rangle$ because the cross term
$\langle {T_\tau{\bf j}^E}^\dag(\tau){\bf j}(0)
\rangle=0$.
We note that the main contribution to 
the thermal current comes from
the effective hopping of the JW fermions
between sites belonging in the same 
sublattice, which is of order~$J^2$.
In zero field where $h'=0$ because $M_z=0$, only 
$L^{22}$ survives, giving the following~contributions:
\begin{equation} 
\begin{array} {lll}
T\kappa_{\rm mag}(\omega) &=&
\int_{RBZ} \frac{dk}{2\pi} \int\frac{d\epsilon}{4\pi}
\frac{n_F(\epsilon)-n_F(\epsilon+\omega)}{\omega}
M_0^2
{\rm Tr}
[{\cal A}(k,\epsilon+\omega)\sigma_0{\cal A}(k,\epsilon)\sigma_0]\cr
&=& 
\int_{RBZ} \frac{dk}{2\pi} \int\frac{d\epsilon}{4\pi} 
\frac{n_F(\epsilon)-n_F(\epsilon+\omega)}{\omega}  
M_0^2 \sum_{p=\pm}\sum_{p'=\pm}\cr
&&\ \ \ \ \ \ \big[\eta^2 - pp'\big(\epsilon+\omega -E_{p'}(k)\big)
\big(\epsilon-E_p(k)\big)\big]
a_{p'}(k,\epsilon+\omega)a_p(k,\epsilon).
\label{spin heat conductivity 3}
\end{array}
\end{equation}

As we did for the magnetic conductivity,
we will derive a semi-classical 
expression for $\kappa_{\rm mag}$.
Hlubek {et al.} \cite{hlubek2010,hlubek2012}
reported that $\kappa_{\rm mag}$ is only limited by extrinsic
scattering processes in the low-$T$ regime. We therefore
use a constant imaginary part for self-energy; i.e., 
${\rm Im}\Sigma=-\eta$, and write for the term 
${\eta} a_p(k,\epsilon)\approx \pi\delta\big(\epsilon-E_p(k)\big)$ 
in the spectral function as was done for the magnetic conductivity.
The result is:
\begin{equation} 
\begin{array} {lll}
T\kappa_{\rm mag}&=&
\sum_{p',p=\pm}\int_{RBZ} \frac{dk}{4\pi}  
M_0^2(k)
\frac{\eta}{[\omega+E_p(k)-E_{p'}(k)]^2+\eta^2} 
\bigg(-\frac{\partial n_F}{\partial \epsilon}\bigg)_{\epsilon=E_p(k)} \cr
&\approx&\frac{1}{2} \sum_{p=\pm}\int_{-\pi}^{\pi} \frac{dk}{4\pi}  
M_0^2(k)
\tau\big(E_p(k)\big)
\bigg(-\frac{\partial n_F}{\partial \epsilon}\bigg)_{\epsilon=E_p(k)},
\label{DC heat conductivity}
\end{array}
\end{equation}
if the main contribution comes from $E_p=E_{p'}$, 
which means that the summation
over index $p'=p$ contributes only one term.
In the low-$T$ limit with $T\ll J$,
Equation~(\ref{DC heat conductivity}) can be evaluated 
yielding $\kappa_{\rm mag}$
linear in temperature. We assume the
scattering processes, represented here by
the scattering rate $\tau(E_p)$, to
be constant.
In the absence of an external magnetic field, 
that is when $h'=0$,
one finds: 
\begin{equation} 
\begin{array} {lll}
\kappa_{\rm mag}&=& \frac{\pi}{3}
\frac{k_B^2}{\hbar}\frac{J_1a}{\hbar}\tau\frac{1}{bc} T \cr
&=& \frac{\pi}{3}
\frac{k_B^2}{\hbar}l_{\rm mag}\frac{1}{bc} T.
%
%
%
\label{DC heat conductivity 2}
\end{array}
\end{equation}

This result is the same as the one found using 
the kinetic estimate in Ref. \cite{hess2007}, namely~$
\kappa_{\rm mag}=\int \frac{dk}{2\pi}c_kv_kl_k,
$
where $c_k = d\epsilon_kn_k/dT$
is the specific heat ($\epsilon_k$ and $n_k$
are the energy and the statistical occupation function 
of the state $k$), $v_k$ the velocity and $l_k$ 
the mean free path of a particle with wavevector~$k$.
In Equation (\ref{DC heat conductivity 2}),
$l_{\rm mag}=\frac{J_1a}{\hbar}\tau$ is the magnetic 
mean-free path, which is assumed to be the same as 
the mean-free path for the magnetic conductivity. 
%
%
%

A magnetic
Wiedemann--Franz law can be defined
as $\frac{\kappa_{\rm mag}}{\sigma_{DC}}=LT$
with $L=\frac{\pi^2k_B^2}{3}$. Here, 
$L$ differs from that in the Wiedemann--Franz law
for true electrons by the absence of the factor $e^2$
in the denominator; $e$ being the electron charge.
When the same self-energy is used for thermal 
and particle transport 
in the Heisenberg antiferromagnets, 
the Wiedemann--Franz law is satisfied, implying that 
both transport phenomena are due to the spin fluctuations represented
here by the motion of the JW fermions.

\section{Discussion and Predictions}

\label{section7}

As far as potential practical applications are concerned, 
we predict that a magneto-motive force (mmf) 
could be realized using a {\it magnetic battery} 
made of a sample of a Heisenberg antiferromagnet in the presence
of a magnetic field with a gradient. Then, a magnetic current
could be generated in a loop connected to this 
magnetic battery, and also made of the
same Heisenberg antiferromagnet in the presence of a uniform 
magnetic field. The magnetic fields involved need not be 
large, and can be chosen to be much smaller than
the saturation field $h_{c}=2J$.  
The magnetic current obviously carries energy,
and can be used in 
spintronics applications.
Note that the magnetic circuits need not be coupled through
interfaces
to the electric circuits providing the magnetic fields. 
If~the predictions of this work 
are confirmed experimentally, then we will have achieved
some sort of practical realization and 
extension in matter of the 
symmetry of Maxwell equations in vacuum
under the duality transformation
${\bf E}\to {\bf B}$ and ${\bf B}\to -{\bf E}$, where ${\bf E}$
and ${\bf B}$ are the electric and magnetic fields, respectively. 
The present proposal for generating magnetic currents 
could be more practical to realize than the
other proposed or used methods that consist of selecting polarized 
spin-up or spin-down electrons.
Because~the~Heisenberg antiferromagnets are insulators,
the magnetic current is not accompanied by charge current at all.
As~mentioned earlier in this work, the spins in the Heisenberg chain 
are polarized along any nonzero applied uniform magnetic 
field.
Using an interface with a metal, 
the~magnetic current in a Heisenberg antiferromagnet 
should in principle give rise 
to an electromotive force by taking advantage of 
the inverse spin Hall effect (ISHE) \cite{saitoh2006}.
In addition, by switching the magnetic current on and off  
in a Heisenberg antiferromagnet, 
the variation in the magnetic field
in these antiferromagnets
may in principle be used to induce an electromotive force
in an ordinary circuit in a contact-less manner contrary to ISHE.

\section{Conclusions}

\label{section8}

In this work, we used the Jordan--Wigner transformation 
to argue in favor of the applicability of 
low-dimensional
Heisenberg antiferromagnets in the area of spintronics. 
We propose that this transformation not only preserves
the spin commutation relations, but reflects also
the duality symmetry that exists between magnetism and electricity
in these materials whose charge (electrons) 
degrees are localized and their
excitations gapped by a large energy. 
The dominating lowest-energy 
excitations are due to the electronic spins (magnetic dipoles),
which are transformed into spinless fermions
by the JW transformation. 
This transformation tells us that the spin magnetic 
moments turn into particles of spin zero.
The spins in these antiferromagnets form
spin liquid states, which are gapless for the Heisenberg 
chain or ladders with an odd number of legs, and gapped for ladders
with an even number of legs. There is 
an interesting similarity between the gapless spin liquid states 
and the Fermi liquid states formed by electrons in conventional metals.

Given that the JW fermions
behave like electrons as far as Fermi statistics is concerned, 
they are convenient for defining and calculating the magnetic current
and magnetic thermal current 
for the spin-$1/2$ Heisenberg antiferromagnets.
The magnetic conductivity and magnetic thermal conductivity, 
calculated in the present work for the Heisenberg chain 
within the bond--mean-field theory,
are found to agree with existing 
results calculated using other methods. 
Finally, the central prediction made here is that 
of generating a magneto-motive force using a Heisenberg
chain-like material in the presence of a magnetic field with a gradient.
We believe that we succeeded to establish a theoretical framework 
for what the experiments in spintronics attempt to do, namely
treating the magnetic degrees of freedom on the same footing
as the electronic ones.

Future work will deal with the Heisenberg ladders
given that several materials of this sort exist in reality, and may be 
of great importance for spintronics.

\vspace{6pt}

\acknowledgments{The author would like to thank A.-M.S. Tremblay 
for his comments on the manuscript.} 



%
%
%
\end{document}